\documentclass[aps,prl,groupedaddress,showpacs
]{revtex4} 
\usepackage{graphicx,graphics,amsfonts,amssymb,color,bm}

\def\be{\begin{equation}} 
\def\ee{\end{equation}} 
\def\ba{\begin{eqnarray}} 
\def\ea{\end{eqnarray}} 
\def\bc{\begin{center}} 
\def\ec{\end{center}} 
\def\sgn{{\rm sgn}} 

\def\p{\partial}
 
\begin{document} 

\title{Quantum Theory of the Third-Harmonic Generation in Graphene} 

\author{S. A. Mikhailov} 
\email[Electronic mail: ]{sergey.mikhailov@physik.uni-augsburg.de} 

\affiliation{Institute of Physics, University of Augsburg, D-86135 Augsburg, Germany} 

\date{\today} 

\begin{abstract}
A quantum theory of the third-harmonic generation in graphene is presented. An analytical formula for the nonlinear conductivity tensor $\sigma^{(3)}_{\alpha\beta\gamma\delta}(\omega,\omega,\omega)$ is derived. Resonant maxima of the third harmonic are shown to exist at low frequencies $\omega\ll E_F/\hbar$, as well as around the frequency $\omega=2E_F/\hbar$, where $E_F$ is the Fermi energy in graphene. At the input power of a CO$_2$ laser ($\lambda\approx 10$ $\mu$m) of about 1 MW/cm$^2$ the output power of the third-harmonic ($\lambda\approx 3.3$ $\mu$m) is expected to be $\simeq 50$  W/cm$^2$. 
\end{abstract} 

\maketitle 

Graphene \cite{Novoselov05,Zhang05}, a one-atom-thin layer of graphite, attracted enormous attention in recent years \cite{Neto09}. In contrast to conventional semiconductors, in which the motion of electrons is governed by the Schr\"odinger equation and the spectrum of electrons is parabolic, $E({\bm p})\propto {\bm p}^2$, the charge carriers in graphene  -- electrons and holes -- obey an effective Dirac equation, have a linear gapless energy dispersion $E({\bm p})\propto \pm |{\bm p}|$, and behave like relativistic massless quasi-particles with the effective ``velocity of light'' $v_F\approx 10^8$ cm/s \cite{Neto09}. These fundamental distinctive features of graphene lead to its unusual electronic and optical properties.

It was predicted \cite{Mikhailov07e} and then experimentally confirmed, both at low (microwave \cite{Dragoman10}) and high (optical \cite{Hendry10}) frequencies, that graphene should demonstrate strongly nonlinear electromagnetic behavior. This nonlinearity directly follows from the linear  energy dispersion of graphene electrons and can be understood from basic physical considerations \cite{Mikhailov07e}. Assume that a particle with the linear spectrum $ E({\bm p})\propto|{\bm p}|$ is placed in the uniform external electric field ${\bm E}(t)={\bm E}_0\cos\omega t$. Then, according to Newton equations of motion the momentum ${\bm p}(t)$ will oscillate as ${\bm p}(t)=-(e{\bm E}_0/\omega)\sin\omega t$ ($e>0$ is the electron charge). In conventional systems with the parabolic electron energy dispersion the velocity, and hence the current, are proportional to the momentum $j(t)\propto v(t)\propto p(t)\propto \sin\omega t$. In contrast, in graphene the velocity ${\bm v}(t)= \p  E/\p {\bm p}\propto{\bm p}(t)/|{\bm p}(t)|$ is a strongly nonlinear function of ${\bm p}(t)$. As a result, the induced current 
\be 
 j(t)\propto v(t)\propto \sgn(\sin\omega t)\propto \sin\omega t+\frac 13 \sin 3\omega t +\dots,
\ee
contains higher frequency harmonics. Just a single graphene layer can thus work as a frequency multiplier which makes it a very interesting material for studying fundamental nonlinear optical processes and may lead to different microwave, terahertz and optoelectronic applications \cite{Mikhailov07e,Mikhailov08a,Mikhailov13c}.

In Refs. \cite{Mikhailov07e,Mikhailov08a} a quasiclassical theory of the nonlinear electromagnetic response of graphene was developed. This theory is based on the solution of the kinetic Boltzmann equation, takes into account only the {\em intra}-band oscillations of the graphene electrons, and is valid at low (microwave/terahertz) frequencies, when $\hbar\omega\lesssim \max\{2|\mu|,T\}$; here $\mu$ is the chemical potential and $T$ is the temperature. At higher (infrared, optical) frequencies the {\em inter}-band electronic transitions should be taken into account, which requires a full quantum nonlinear-response theory. In this paper we present such a theory. We consider a graphene layer lying in the plane $z=0$ under the action of a uniform ac electric field $E_\alpha(t)= E_\alpha^0e^{-i\omega t} + \textrm{c.c.}$ and calculate the third-order conductivity tensor defined as 
\be
j_\alpha^{(3)}(t)= \sigma^{(3)}_{\alpha\beta\gamma\delta} (\omega,\omega,\omega)
E_\beta^0E_\gamma^0E_\delta^0e^{-i3\omega t} + \textrm{c.c.},
\label{nonlin3current}
\ee
where $j_\alpha^{(3)}(t)$ is the induced third-harmonic current, and c.c. means the complex conjugate. The results obtained take into account both the intra- and inter-band quantum transitions and describe the third-harmonic response of graphene at all frequencies from radiowaves up to visible light. 

The spectrum of electrons ($l=2$) and holes ($l=1$) in graphene can be described by the tight-binding Hamiltonian $\hat H_0$ with the eigen-energies 
\be
E_{l{\bm k}}=(-1)^lt \left|{\cal S}_{\bm k}\right|,\ \ 
{\cal S}_{\bm k}=1+2\cos(k_xa/2)e^{i\sqrt{3}k_ya/2},
\label{energy}
\ee
and the eigen-functions $|\lambda\rangle\equiv|l{\bm k}\sigma\rangle$; here ${\bm k}$ is the electron wave-vector, $\sigma$ is the spin quantum number, $a$ is the graphene lattice constant, and $t$ is the tight-binding transfer integral. To calculate the system response we solve the quantum kinetic equation $i\hbar\p\hat\rho/\p t=[\hat H_0-e\phi({\bm r},t),\hat\rho]$ for the density matrix $\hat\rho$. The electric potential here,
\be 
 \phi({\bm r},t)= \phi_{{\bm q}\omega}e^{i{\bm q}\cdot {\bm r}-i\omega t+\gamma t}+\textrm{c.c.},\ \ \gamma\to+0,
\label{extpotent}
\ee
determines the electric field ${\bm E}({\bm r},t)= -i{\bm q}\phi_{{\bm q}\omega}e^{i{\bm q}\cdot {\bm r}-i\omega t+\gamma t}+\textrm{c.c.}$ and is assumed to be (for a moment) space-dependent. The limit ${\bm q}\to{\bm 0}$ will be taken later on; this should be done with care since the terms linear in ${\bm q}$ must be kept. 

\begin{figure}
\includegraphics[width=8.5cm]{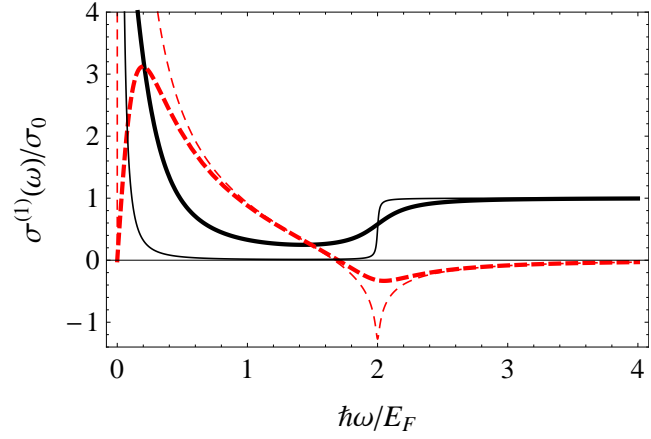}
\caption{\label{fig:sigma(1)} The real (black solid) and imaginary (red dashed curve) parts of the first-order conductivity $\sigma^{(1)}(\omega)$ as a function of the frequency $\hbar\omega/E_F$ at $\Gamma=\hbar\gamma/E_F=0.2$ (thick curves) and $\Gamma=0.01$ (thin curves); $\sigma_0=e^2/4\hbar$ is the universal optical conductivity.  }
\end{figure}

First, consider the linear response. Then the $({\bm q}\omega)$-Fourier component of the current reads
\be 
j^{(1),\alpha}_{{\bm q}\omega}=
\frac {e^2}{2S}\phi_{{\bm q}\omega}
\sum_{\lambda\lambda'} 
\langle\lambda'|\{\hat v_\alpha,e^{-i{\bm q}\cdot {\bm r}}\}_+|\lambda\rangle 
\frac {f_{\lambda'}-f_{\lambda}}{E_{\lambda'}-E_{\lambda}+\hbar(\omega + i\gamma)} 
\langle\lambda|e^{i{\bm q}\cdot {\bm r}}|\lambda'\rangle ,
\label{current1}
\ee
where $S$ is the area of the sample, $\hat v_\alpha$ is the velocity operator, and $\{\dots\}_+$ means the anti-commutator. At small ${\bm q}$ (in the linear order) the matrix element of the function $e^{i{\bm q}\cdot {\bm r}}$ assumes the form 
\be 
\langle\lambda|e^{i{\bm q}\cdot {\bm r}}|\lambda'\rangle\approx
\delta_{\sigma\sigma'}
\left(\delta_{l'l} \delta_{{\bf k},{\bf k'+q}} 
 +(1-\delta_{l'l})\delta_{{\bf k},{\bf k'}} 
\frac {1}2
q_\alpha \zeta^\star_{\bf k} \frac{\p \zeta_{{\bf k}}}{\p  k_\alpha}
\right),
\label{MEexp}
\ee
where $\zeta_{\bf k}={\cal S}_{\bf k}/|{\cal S}_{\bf k}|$. The first and second terms in parenthesis here correspond to the intra-band ($l=l'$) and inter-band ($l\neq l'$) contributions, respectively. Substituting the matrix element (\ref{MEexp}) in Eq. (\ref{current1}), taking the limit $q\to0$ in the rest of the formula, and calculating the integrals over $d{\bm k}$ at $T=0$ (we assume that $T\ll|\mu|=E_F$), we obtain the first-order conductivity \cite{Gusynin06b,Falkovsky07a,Mikhailov07d}:
\be
\sigma_{\alpha\beta}^{(1)}(\omega)=
\delta_{\alpha\beta} 
\frac{ie^2g_sg_v}{4\pi\hbar}
\Bigg(\frac 1{\Omega+i\Gamma} + 
\frac {1}{4}
\ln\frac{2-(\Omega + i \Gamma)}{2+(\Omega + i \Gamma)}
\Bigg).
\label{linconduct}
\ee
Here $\Omega=\hbar\omega/|\mu|$, $\Gamma=\hbar\gamma/|\mu|$, and $g_s=g_v=2$ are the spin and valley degeneracies. The first and second terms in (\ref{linconduct}) are the intra-band (Drude) and inter-band conductivities, respectively. The quantity $\gamma$ can be treated as a phenomenological scattering parameter which accounts for the broadening of resonances. The logarithm in (\ref{linconduct}) is a complex-valued function, which acquires an imaginary part at $|\Omega|\gtrsim 2$ and leads to the universal optical conductivity $\sigma^{(1)}(\omega)=\sigma_0=e^2/4\hbar$ at large frequencies \cite{Nair08}. 
Figure \ref{fig:sigma(1)} shows that thus calculated conductivity $\sigma^{(1)}(\omega)$ is in good agreement with experiments (compare, e.g., with Fig. 2 in Ref. \cite{Li08}). 

In the third order a similar calculation gives the following $(3{\bm q},3\omega)$-Fourier component of the current
\ba 
j_{3{\bm q},3\omega}^{(3),\alpha}&=&
\frac {e^4}{2S}
(\phi_{{\bm q}\omega})^3
\sum_{\lambda\lambda'\lambda''\lambda'''} 
\frac { \langle\lambda'| \{\hat v_\alpha, e^{-i3{\bm  q}\cdot{\bm r}}\}_+|\lambda\rangle
\langle\lambda| e^{i{\bm q}\cdot{\bm r}}|\lambda'''\rangle
\langle\lambda'''|e^{i{\bm q}\cdot{\bm r}} |\lambda''\rangle 
\langle\lambda''| e^{i{\bm q}\cdot{\bm r}}  |\lambda'\rangle 
}
{E_{\lambda'}-E_{\lambda}+3\hbar(\omega +i\gamma)}
\nonumber \\ &\times&
\left[
\frac{1}
{E_{\lambda'}-E_{\lambda'''} + 2\hbar (\omega +i\gamma) }
\left(
\frac {f_{\lambda'}-f_{\lambda''}}
{E_{\lambda'}-E_{\lambda''}+\hbar(\omega+i\gamma)} 
-
\frac {f_{\lambda''}-f_{\lambda'''}}
{E_{\lambda''}-E_{\lambda'''}+\hbar(\omega+i\gamma)} 
\right)\right.
\nonumber \\  &-& \left.
\frac{1}
{E_{\lambda''}-E_{\lambda} + 2\hbar (\omega +i\gamma) }
\left(
\frac {f_{\lambda''}-f_{\lambda'''}}
{E_{\lambda''}-E_{\lambda'''}+\hbar(\omega+i\gamma)} 
-
\frac {f_{\lambda'''}-f_{\lambda}}
{E_{\lambda'''}-E_{\lambda}+\hbar(\omega+i\gamma)} 
\right)
\right] .
\ea
Now we have a product of three matrix elements of the type (\ref{MEexp}), 
\be 
\langle\lambda| e^{i{\bm q}\cdot{\bm r}}|\lambda'''\rangle
\langle\lambda'''|e^{i{\bm q}\cdot{\bm r}} |\lambda''\rangle 
\langle\lambda''| e^{i{\bm q}\cdot{\bm r}}  |\lambda'\rangle,\label{ME}
\ee 
each being the sum of the intra-band and inter-band contributions. Expanding this product we get altogether eight terms; only one of them (proportional to $\delta_{ll'''}\delta_{l'''l''}\delta_{l''l'}$) corresponds to the purely classical (intra-band) contribution found previously  \cite{Mikhailov07e,Mikhailov08a}. 
Calculating now all eight terms we get 
\be 
\sigma^{(3)}_{\alpha\beta\gamma\delta} (\omega,\omega,\omega)=
i \sigma^{(3)}_0
S_{\alpha\beta\gamma\delta}(\Omega,\Gamma),
\label{sigma3result}
\ee
where 
\be 
\sigma^{(3)}_0=\frac{e^4g_sg_v\hbar v_F^2}{16 \pi E_F^4},\label{pref}
\ee 
and 
\ba 
S_{\alpha\beta\gamma\delta}(\Omega,\Gamma)&=&
\frac {16\delta_{\alpha\delta}\delta_{\beta\gamma}  
} {(\Omega+i\Gamma)[4-(\Omega+i\Gamma)^2]^2}
- \frac {2\delta_{\alpha\beta}\delta_{\gamma\delta} }
{ (\Omega  +i\Gamma) [4-(\Omega+i\Gamma)^2]} 
\nonumber \\ &+& \Big(\delta_{\alpha\beta} 
 \delta_{\gamma\delta}   
+ \delta_{\alpha\gamma} 
\delta_{\beta\delta}   +
 \delta_{\alpha\delta} 
\delta_{\beta\gamma}\Big)
\Bigg[
\frac 1{3(\Omega +i\Gamma)^3}-\frac{(\Omega+i\Gamma)}{[4-(\Omega+i\Gamma)^2]^2}
\nonumber \\ &+&
\frac{3}{16(\Omega +i\Gamma)^4}
\Bigg(
\ln\frac{2-(\Omega +i\Gamma)}{2+(\Omega +i\Gamma)}
 -
\frac {1}{27}
\ln
\frac{2-3(\Omega +i\Gamma)}
{2+3(\Omega +i\Gamma)}
\Bigg)
\Bigg].\label{Sf}
\ea
The tensor $\sigma^{(3)}_{\alpha\beta\gamma\delta} (\omega,\omega,\omega)$ satisfies the relation
$
\left(\sigma^{(3)}_{\alpha\beta\gamma\delta} (\omega,\omega,\omega)\right)^\star=\sigma^{(3)}_{\alpha\beta\gamma\delta} (-\omega,-\omega,-\omega),
$ 
where the star means the complex conjugate; its non-zero components are $\sigma^{(3)}_{xxyy}=\sigma^{(3)}_{yyxx}$, $\sigma^{(3)}_{xyxy}=\sigma^{(3)}_{yxyx}$, $\sigma^{(3)}_{xyyx}=\sigma^{(3)}_{yxxy}$, and $\sigma^{(3)}_{xxxx}=\sigma^{(3)}_{yyyy}=
\sigma^{(3)}_{xxyy}+
\sigma^{(3)}_{xyxy}+
\sigma^{(3)}_{xyyx}$. 

\begin{figure}
\includegraphics[width=8.5cm]{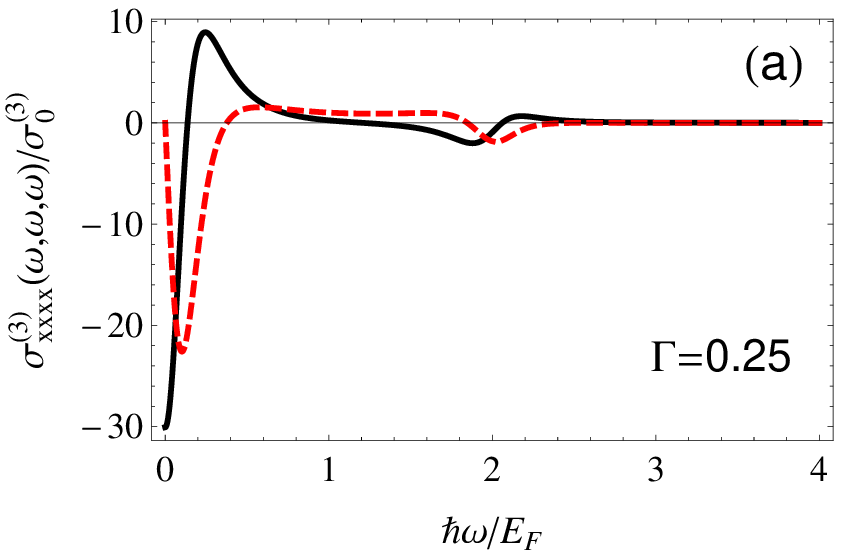}\hfill
\includegraphics[width=8.5cm]{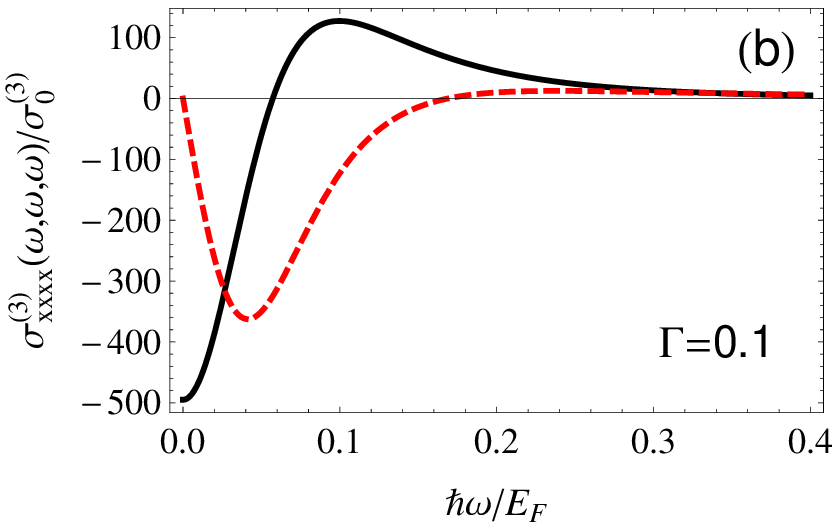}\\
\includegraphics[width=8.5cm]{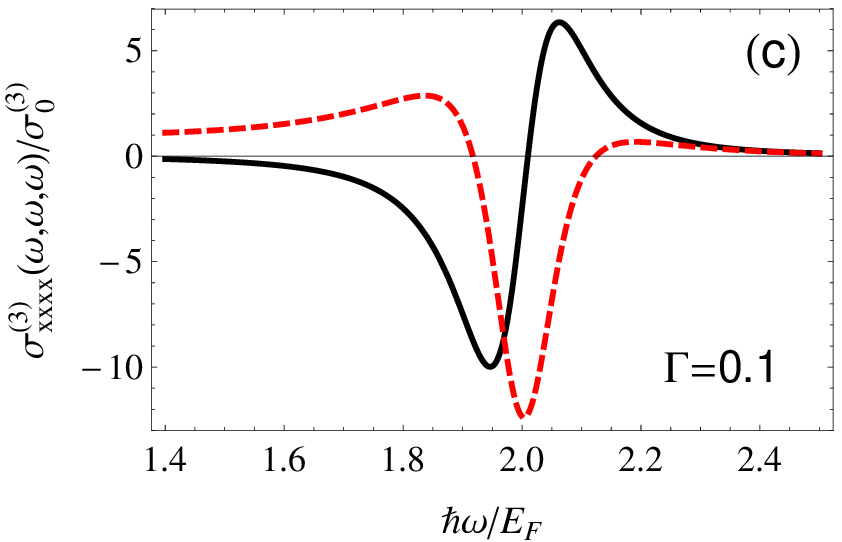}\hfill
\includegraphics[width=8.5cm]{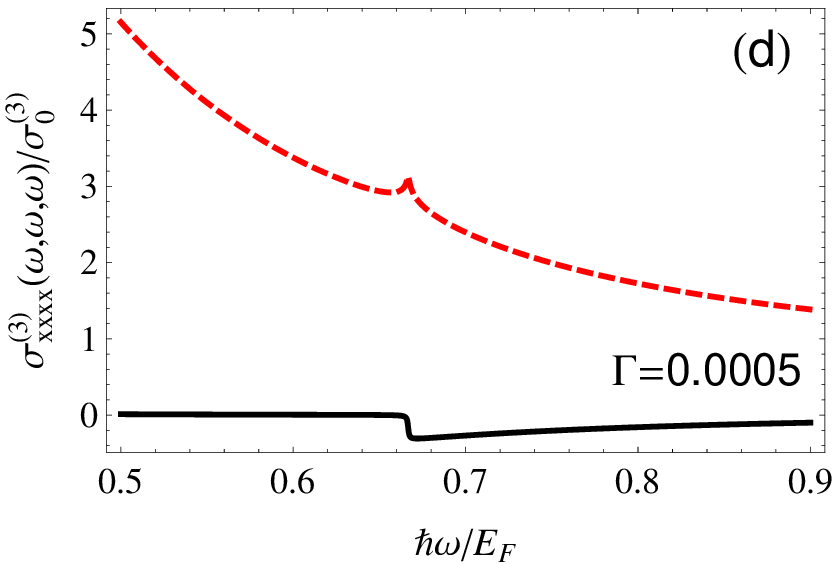}
\caption{\label{fig:sigma3} The real (black solid) and imaginary (red dashed curve) parts of the third-order conductivity $\sigma^{(3)}_{xxxx}(\omega,\omega,\omega)$ as a function of the frequency $\hbar\omega/E_F$: (a) an overview in a broad frequency range at $\Gamma=0.25$; a detailed view at (b) low (microwave, terahertz, $\Omega\ll 1$) and (c) high (infrared, optical, $\Omega\simeq 2$) frequencies, $\Gamma=0.1$; (d) the weak logarithmic feature around $\Omega=2/3$ at $\Gamma\ll 1$. 
}
\end{figure}

If the external electric field is linearly polarized, the third-harmonic response is determined by the function $\sigma^{(3)}_{xxxx} (\omega,\omega,\omega)$. Figure \ref{fig:sigma3} shows the frequency dependence of $\sigma^{(3)}_{xxxx}$ in different frequency ranges. In general, there are two resonant features, Fig. \ref{fig:sigma3}(a). The low-frequency resonance ($\Omega\ll 1$) is mainly due to the classical (intra-band) contribution and is larger than the high-frequency resonance at $\Omega\simeq 2$. When the scattering parameter $\Gamma$ decreases, Fig. \ref{fig:sigma3}(b),(c), the amplitudes of both resonances, as well as the difference between them, dramatically grow (notice the difference of the vertical axis scales in different plots). The largest contribution to the high-frequency resonance at $\Omega\simeq 2$ is provided by the terms in (\ref{ME}) containing two intra-band and one inter-band factors. The logarithmic feature at $\Omega\simeq 2/3$ (see the last term in Eq. (\ref{Sf})) can be seen only at extremely small values of $\Gamma$, Fig. \ref{fig:sigma3}(d).

\begin{figure}
\includegraphics[width=8.5cm]{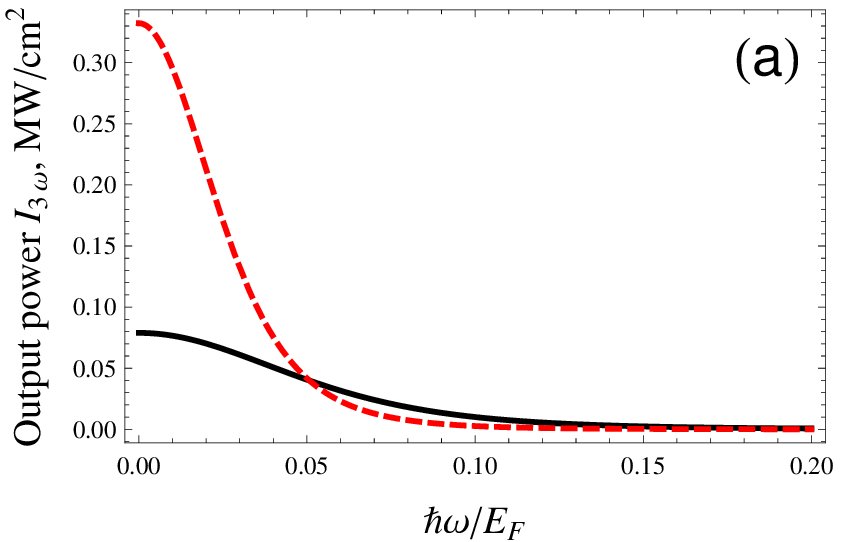}
\includegraphics[width=8.5cm]{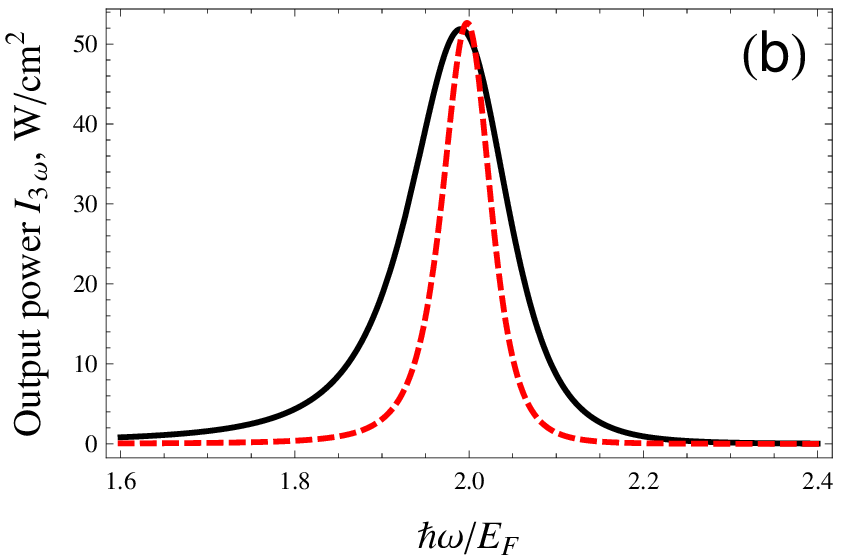}
\caption{\label{fig:power} The output power (\ref{I3w}) of the third-harmonic wave in the regime of (a) low ($\Omega\ll 1$) and (b) high ($\Omega\simeq 2$) frequencies; black solid curves: $\Gamma=0.1$, $n_s=3\times 10^{11}$ cm$^{-2}$,  $I_\omega=1$ MW/cm$^{-2}$; red dashed curves: $\Gamma=0.05$, $n_s= 10^{12}$ cm$^{-2}$, $I_\omega=2$ MW/cm$^{-2}$. The frequency of the incident wave corresponding to the resonant condition $\hbar\omega=2E_F$ are about 31 THz (the wavelength $\lambda=9.7$ $\mu$m) for the black curve and 56.4 THz (the wavelength $\lambda=5.3$ $\mu$m) for the red curve. The corresponding third-harmonic frequency/wavelength are 92.7 THz (3.24 $\mu$m) and 170 THz (1.77 $\mu$m), respectively. 
}
\end{figure}

The absolute value of the emitted third-harmonic intensity $I_{3\omega}$ can be calculated using the prefactor (\ref{pref}). For example, for a single suspended graphene layer in free space we get
\be
I_{3\omega}=I_\omega
\left(
 \frac {  \alpha^2 I_\omega}{ \pi  n_s^2 \hbar v_F^2}
\left|S_{xxxx}(\Omega,\Gamma)
\right| \right)^2,\label{I3w}
\ee
where $\alpha=e^2/\hbar c$ is the fine structure constant, $n_s$ is the electron (or hole) density, and $I_\omega$ is the intensity of the incident (linearly polarized) wave with the frequency $\omega$. Figure \ref{fig:power} shows that at $I_\omega\simeq 1-2$ MW/cm$^2$ the output third-harmonic intensity can be as large as $\sim 0.3$ MW/cm$^2$ at low frequencies and $\sim 50$ W/cm$^2$ near the high-frequency resonance $\Omega\simeq 2$. It should, however, be noticed that in our theory the graphene-layer size is assumed to be infinite, in practical terms, much larger than the wavelength of radiation (several mm at terahertz frequencies and several micron at the visible-light frequencies). If this condition is not satisfied (e.g. at radio- or microwave frequencies) the output power will be accordingly smaller. 

I am grateful to Michael Glazov for useful discussions and the Deutsche Forschungsgemeinschaft for the financial support of this work.


\end{document}